\documentclass{article}

\usepackage{microtype}
\usepackage{graphicx}
\usepackage{subfigure}
\usepackage{enumitem}
\usepackage{booktabs} 

\usepackage{hyperref}
\usepackage{multirow}
\usepackage{algpseudocodex}

\definecolor{comm}{gray}{0.5}
\usepackage{algorithm}

\usepackage[accepted]{icml2024}

\usepackage{amsmath}
\usepackage{amssymb}
\usepackage{mathtools}
\usepackage{amsthm}
\usepackage{float}

\usepackage[capitalize]{cleveref}
\crefname{section}{Sec.}{Secs.}

\theoremstyle{plain}

\theoremstyle{definition}

\theoremstyle{remark}

\newcommand{\method}[0]{\textsc{ReGAL}}

\newcommand{\interval}[1]{_{\pm\text{ #1}}}
\newcommand{\fname}[1]{\texttt{\textbf{#1}}}
\newcommand{\sfname}[1]{$\tt{#1}$}
\newcommand{\sparagraph}[1]{\textbf{#1 \hspace{0.3em}}}

\definecolor{deepblue}{rgb}{0,0,0.5}
\definecolor{deepred}{rgb}{0.6,0,0}
\definecolor{deepgreen}{rgb}{0,0.5,0}

\DeclareFixedFont{\ttb}{T1}{txtt}{bx}{n}{8} 
\DeclareFixedFont{\ttm}{T1}{txtt}{m}{n}{8}  

\usepackage{listings}

\lstnewenvironment{codeexample}
   {\lstset{escapechar=`,linewidth=\linewidth}}
   {}
\lstalias{none}{}
\lstnewenvironment{prompt}
   {\lstset{escapechar=~,linewidth=\textwidth, language=none}}
   {}

\newcommand\pythonstyle{\lstset{
language=Python,
basicstyle=\ttm,
morekeywords={self},              
keywordstyle=\ttb\color{deepgreen},
emph={MyClass,__init__,calculate_midpoint,calculate_square,calculate_total_cost,add_prices,craft_object_with_ingredients,check_and_get_object,get_date_today,get_date_one_week_from_today,get_date_one_week_ago,get_date_one_year_ago,draw_small_5gon,draw_semicircle},          
emphstyle=\ttb\color{deepblue},    
stringstyle=\color{deepgreen},
frame=none,                         
showstringspaces=false
}}

\lstnewenvironment{python}[1][]
{
\pythonstyle
\lstset{#1}
}
{}

\usepackage[textsize=tiny]{todonotes}

\icmltitlerunning{\method{}: Refactoring Programs to Discover Generalizable Abstractions}

\begin{document}

\twocolumn[
\icmltitle{\method{}: Refactoring Programs to Discover Generalizable Abstractions}

\icmlsetsymbol{equal}{*}

\begin{icmlauthorlist}
\icmlauthor{Elias Stengel-Eskin}{equal,yyy}
\icmlauthor{Archiki Prasad}{equal,yyy}
\icmlauthor{Mohit Bansal}{yyy}
\end{icmlauthorlist}

\icmlaffiliation{yyy}{UNC Chapel Hill} 

\icmlcorrespondingauthor{Elias Stengel-Eskin}{esteng@cs.unc.edu}

\icmlkeywords{Program induction, code refactoring, tool discovery, LLMs}

\vskip 0.3in
]

\printAffiliationsAndNotice{\icmlEqualContribution} 

\begin{abstract}
While large language models (LLMs) are increasingly being used for program synthesis, they lack the global view needed to develop useful abstractions; they generally predict programs one at a time, often repeating the same functionality. 
Generating redundant code from scratch is both inefficient and error-prone. 
To address this, we propose \textbf{Re}factoring for \textbf{G}eneralizable \textbf{A}bstraction \textbf{L}earning (\method{}), a gradient-free method for learning a library of \emph{reusable} functions via code \emph{refactorization}, i.e., restructuring code without changing its execution output. 
\method{} learns from a small set of existing programs, iteratively verifying and refining its abstractions via execution.
We find that the shared function libraries discovered by \method{} make programs \emph{easier to predict} across diverse domains.
On five datasets -- LOGO graphics generation, Date reasoning, TextCraft (a Minecraft-based text-game) MATH, and TabMWP -- both open-source and proprietary LLMs improve in accuracy when predicting programs with \method{} functions.
For CodeLlama-13B, \method{} results in absolute accuracy increases of $11.5\%$ on LOGO, $26.1\%$ on date understanding, and $8.1\%$ on TextCraft, outperforming GPT-3.5 in two of three domains.   
Our analysis reveals \method{}'s abstractions encapsulate frequently-used subroutines as well as environment dynamics.\footnote{Code: \url{https://github.com/esteng/regal_program_learning}.}
\end{abstract}

\section{Introduction} \label{sec:intro}
\begin{figure}[ht]
    \centering
    \includegraphics[trim={0.3cm 0.1cm 0.2cm 0cm},clip,scale=0.375]{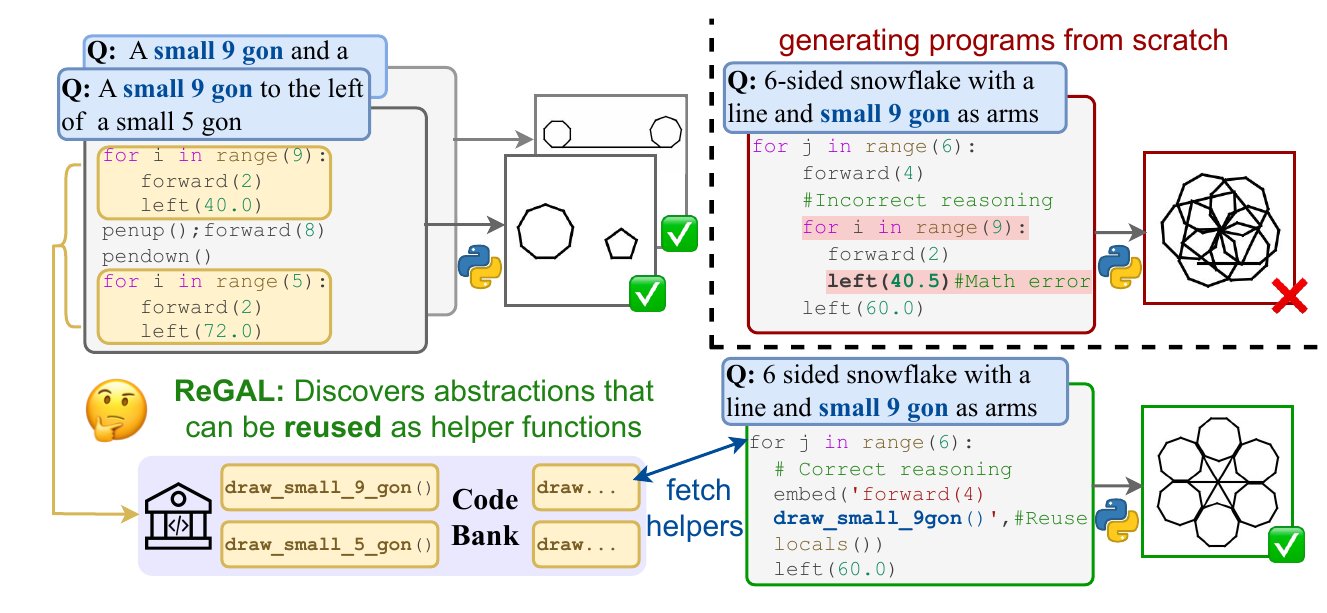}
    \vspace{-2.25em}
    \caption{\method{} trains by refactoring primitive-only programs into abstractions that are verified and stored. 
    These abstractions have two benefits: \textbf{Reusability}: Rewriting the same code multiple times leads to errors; 
    \textbf{Abstraction}: \method{} makes prediction easier by allowing matching between the query and the abstractions.  
    }
    \label{fig:fig1}
\end{figure}
An increasing range of tasks can be tackled by using a large language model (LLM) to generate an executable program for a given query;
this paradigm has been applied in computer vision \citep{suris.d.2023, gupta.s.2018, cho.j.2023vpeval}, robotics \citep{ahn.m.2022saycan, singh.i.2023}, tool use \citep{schick.t.2022, lu2023chameleon, qin2023toolllm}, and complex reasoning \citep{lyu.q.2023faithful}.
In all these cases, the overall program generation framework is the same: an individual query is given (along with an instructive prompt) to an LLM, which produces a program that, when executed, yields the desired result. 
Crucially, each program is generated \textit{independently} (as shown in \cref{fig:fig1}), with no reference to other queries or programs, and is composed of  \emph{primitive} operations, i.e., the domain language's built-in operations.
This approach has two major and related limitations: 

\textbf{1) Lack of Reusability}: Each program is designed as a one-off script to solve a given example but is not reused by other examples. 
This increases redundancy and can result in unnecessary errors: for two examples requiring a shared subroutine, the model might correctly generate the subroutine in one example and make a mistake in the other. 
For instance, in \cref{fig:fig1} (top) although the ``primitive-only'' model had previously generated nonagons, it draws a polygon with an incorrect angle.
\method{}'s \sfname{draw\_small\_9gon\text{()}} function, on the other hand, executes correctly.
    
\textbf{2) Lack of Abstraction}: Shared abstractions can improve accuracy by making skills more accessible to the model. 
When generating from primitives alone, the model must interpret the query and generate the correct mapping from the query to \emph{multiple primitives}, requiring more reasoning.
The model's overall task becomes \emph{easier} when it uses interpretable abstractions, as it is choosing a function name from a library instead of reasoning from scratch. 
In \cref{fig:fig1} (bottom) a model augmented with abstractions can match the sub-query \emph{``a small 9 gon''} to \sfname{draw\_small\_9gon\text{()}}; with this part of the task simplified, the model reasons correctly about the remaining code, while the primitive-only model fails to correctly embed the shape in a loop.

Both limitations can be traced to a \emph{lack of global context} as the model sees each example separately, so it lacks a mechanism for developing reusable abstractions.
This differs greatly from how humans write code: 
generally, developers might start solving individual tasks with one-off solutions, but quickly begin to develop a library of shared abstractions and code snippets for related problems, thereby reducing redundancy in their code, promoting efficiency and readability \citep{mcconnell.s.2004code, downey.a.2012}.
Furthermore, functions can be \emph{verified}: once we have tested a function, we can rely on it in the future -- something that is harder to do for ever-changing one-off code snippets. 
Such abstraction and verification is only sensible if the code synthesis process takes place over the course of multiple examples. 
In other words, if presented with a single, one-off task, there is no reason not to write a one-off script. 

While abstraction offers numerous benefits, it comes with the risk of over-fitting, where a function tailored to a specific example loses its generalizability. For instance, in \cref{fig:fig1}, a function like \sfname{draw\_9gon\_snowflake\text{()}} may perfectly match one example but fails to generalize. Conversely, \sfname{draw\_small\_9gon\text{()}} is a more versatile function applicable in various contexts. The ability to produce novel programs using primitive operations needs to be balanced with the benefits of encoding subroutines into reusable abstractions \citep{odonnell.t.2015}. A similar balance between flexibility and efficiency appears in a variety of domains, including language \citep{odonnell.t.2015, yang.c.2016price}, biology \citep{futuyma.d.1988evolution}, manufacturing \citep{flynn.b.1987applications}, and programming \citep{ellis.k.2021}. To strike this balance in LLM-based program synthesis, we propose \textbf{Re}factoring for \textbf{G}eneralizable \textbf{A}bstraction \textbf{L}earning (\method{}). \method{} refines abstractions iteratively by refactoring programs as well as verifying, correcting, and pruning abstractions such that overly specific or incorrect programs are improved upon or removed.  \method{} relies on two key elements: a small set of programs using primitive operations (i.e., \emph{primitive programs}) and an execution environment (e.g., Python). Importantly, we show \method{} can learn from LLM-generated programs without requiring any human annotations.

\method{} follows a familiar train-test paradigm: during \method{}'s modular training phase (see \cref{fig:method}), it iteratively refactors a small set of \emph{(query, program)} examples to produce a library of useful abstractions.
\method{} uses an LLM to write helper functions for a batch of examples, which are verified against the expected result; successful helper functions are added to the library and the refactored program serves as an example of the function's usage. 
\method{} can take success feedback into account to correct and debug errors, and it periodically edits the helper functions to make them more generalizable or -- if they cannot be made more generic -- prunes functions that are overly specific. 
Note that the training is gradient-free, relying on a frozen LLM to refactor programs.
In the testing phase, an LLM agent is tasked with predicting programs for test queries.
The agent has access to \method{}'s library of helper functions and demonstrations of how to use them. 

We demonstrate the broad applicability of \method{} by testing it on five diverse datasets: LOGO \citep{ellis.k.2021, wong.l.2021}, a program induction task; a date reasoning task \citep{srivastava.a.2022bigbench} known to challenge LLMs \citep{suzgun.m.2022challenging}; TextCraft \citep{prasad2023adapt}, a text-based game for crafting Minecraft objects; a subset of MATH \citep{hendrycks2021measuring} which contains algebra word problems, and TabMWP \citep{lu2022dynamic}, a collection of math quesitons about tables. Across these tasks, \method{} significantly improves the accuracy of the predicted programs from various LLMs -- especially open-source LLMs -- over a baseline that predicts primitive programs (i.e., programs without \method{}'s abstractions). For instance, CodeLlama-13B's \citep{roziere.b.2023codellama} accuracy increases by $11.5\%$, $26.1\%$, and $8.1\%$ on LOGO, Date, and TextCraft respectively, surpassing larger models like GPT-3.5 (cf. \cref{sec:results}). 
In \cref{sec:analysis}, we show that \method{}'s abstractions are reusable across examples, encapsulate key domain functionalities, and we include an error analysis further highlighting the features that make \method{} effective. 
\cref{sec:efficiency} reveals that \method{} can improve over baseline primitive programs with minimal examples, yielding major improvements even with a $50\%$ reduced training set. 

\vspace{-0.5em}
\section{Related Work}
\vspace{-0.5em}
\sparagraph{Program Induction.}
Program induction involves learning a symbolic and programmatic mapping of inputs to outputs. 
Humans are adept at this kind of ``rule-learning''~\citep{marcus1999rule, furnkranz2012foundations}.
\method{} also aims to learn a set of general functions that can be used to map inputs to outputs, i.e., a form of program induction. 
\citet{ellis.k.2021} present DreamCoder, a method for combining program induction and synthesis that uses a wake-sleep Bayesian learning method to learn programs.
\citet{wong.l.2021} extend this work to incorporate language, using an alignment model as part of the joint model. 
Like \citet{ellis.k.2021}, \citet{grand.g.2023} adopt a similar symbolic search procedure, but use an LLM to document abstractions. 
The symbolic search procedure used by this line of past work has relied on data structures that assume the domain language is $\lambda$-calculus \citep{lake.b.2015human, ellis.k.2021, wong.l.2021, grand.g.2023}, which is not typically used for software development.
In contrast, \method{} has an LLM-based search procedure, allowing us to use flexible languages like Python, which are more commonly used by developers and better represented in pre-training data. 

\sparagraph{Program Synthesis and Tool Use.}
Tool use by LLMs \citep{schick.t.2022, mialon2023augmented} refers to a form of program synthesis or semantic parsing where an LLM generates API calls to external tools (e.g., calculators, search functions, etc.). 
This formulation has also been applied to reasoning tasks \citep{lyu.q.2023faithful, chen.w.2022program} as well as other domains such as computer vision \citep{suris.d.2023, gupta.t.2023visual, cho.j.2023vpeval}, summarization \citep{saha.s.2022summarization}, and robotics \citep{ahn.m.2022saycan, singh.i.2023, huang.w.2022, huang.w.2023voxposer}. 
Past works have attempted to induce tools from examples.
\citet{cai.t.2023} induce tools using an LLM for reasoning tasks from BigBench \citep{srivastava.a.2022bigbench}; unlike our work, their system generates one tool per task. 
While this can offer benefits for homogenous reasoning tasks (e.g., sorting words alphabetically), heterogenous tasks like the ones we explore require multiple functions. 
More akin to our work, \citet{yuan.l.2023craft}, \citet{qian2023creator}, and \citet{wang2024trove} induce multiple tools for vision and math tasks using an LLM-based framework which also includes retrieval-based parsing.
In addition to focusing on different domains, we place an emphasis on learning a shared code bank that can be used by multiple LLMs to generate programs including several open-source LLMs.
We also differ in our focus on refactoring, and in the amount of information we provide to the refactoring model: unlike \citet{yuan.l.2023craft} and \citet{qian2023creator}, we do not provide in-context examples of the kinds of tools we want the model to create, investigating instead what abstractions the model builds without domain-specific guidance. 
While \citet{wang2024trove} also prunes their learned functions, their method does not involve refactoring; in contrast, we learn functions by refactoring \emph{and} periodically edit the codebank (in addition to pruning it).

\sparagraph{Induction in Interactive Domains.}
\citet{wang.g.2023voyager} also induce functions in a Minecraft domain; however, theirs are written and stored based on one iteration.
In contrast, our work refactors programs in a group and tests and refines them across the training process, showing generalization in multiple domains.
Other prior work learns a library of abstractions for planning in embodied domains~\citep{wong.l.2023learning, majumder2023clin}. 
While we share a similar motivation, \method{} operates in the space of generating executable programs instead of PDDL operators~\citep{wong.l.2023learning} or causal textual feedback~\citep{majumder2023clin}.
Similarly, our work aligns with prior efforts in LLM-based task decomposition \citep{khot2023decomposed, prasad2023adapt}, where skills are reused across multiple task instances. 
However, these approaches manually identify atomic skills and require the LLM to repeatedly execute skills from scratch. In contrast, \method{} provides a way of automatically discovering such abstractions and reusing them via helper functions.

\vspace{-0.5em}
\section{Methodology}
\begin{figure*}[ht]
    \centering
    \includegraphics[width=0.925\linewidth]{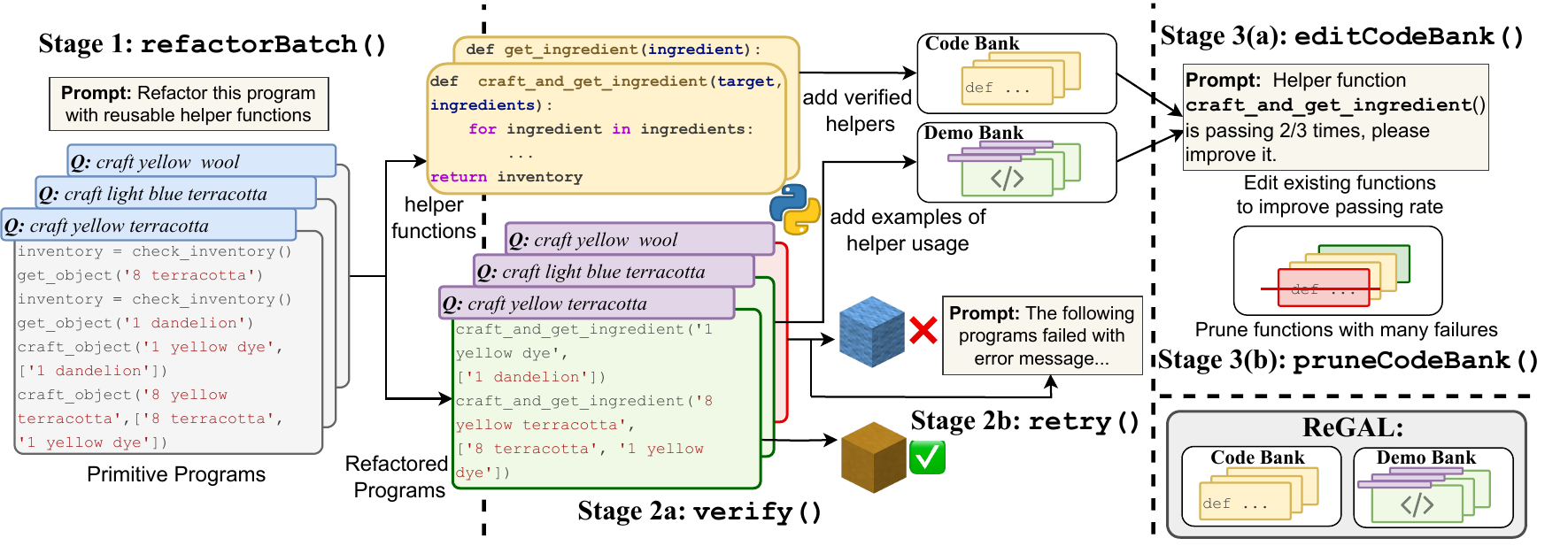} 
    \vspace{-1em}
    \caption{\method{} starts by refactoring a batch of primitive programs to develop a set of modified programs and helper functions (\textbf{Stage 1}). 
    It then verifies the results of refactored programs, optionally retrying failed programs according to environment feedback. 
    Useful helper functions are added to the Code Bank along with example usage added to the Demo Bank (\textbf{Stage 2}). 
    Periodically, we edit and prune the Code Bank to improve its functions (\textbf{Stage 3}). 
    At test time, the \method{} agent has access to the Code Bank, the Demo Bank, and the remaining original programs. 
    It is compared against a baseline agent which has access to a larger number of original programs. }
    \label{fig:method}
\end{figure*}

\vspace{-0.5em}
In this section, we describe the overall pipeline of our method: \textbf{Re}factoring for \textbf{G}eneralizable \textbf{A}bstraction \textbf{L}earning  (\method{}). 
\method{} consists of two phases: the \emph{training} or induction stage where abstractions (i.e., helper functions) are learned, and the \emph{testing} or synthesis stage, where abstractions are used to generate programs for test queries.
During training, as illustrated in \cref{fig:method}, \method{} discovers reusable abstractions by generating candidate helper functions, validating their correctness, and debugging via editing and pruning of ineffective helper functions.
Given a set of demonstrations $(q, p)$ of queries $q$ and gold primitive programs $p$, we first preprocess the data to cluster examples based on query similarity, described in \cref{sec:method_pre}. 
The training stage then builds abstractions by refactoring primitive programs in batches (\cref{sec:method_train}), while the testing stage solves new queries by generating programs that glue together the learned abstractions with primitive operations (\cref{sec:method_test}). 
We use GPT-3.5 for training; at test time we use a range of LLMs, focusing on freely available open-source LLMs. 

\vspace{-0.5em}
\subsection{Preprocessing} \label{sec:method_pre}
\vspace{-0.5em}
Before training, we preprocess queries and programs $(q, p)$ by (optionally) adding comments, clustering them into related batches, and sorting them by approximate difficulty.

\sparagraph{Adding Comments.} We optionally add comments to align the query with the primitive program, enabling the model to generate the correct abstractions. 
We present each $(q,p)$ pair to GPT-3.5 independently, with a prompt asking the model to comment $p$ based on $q$; we then verify that the commented code executes to the same result. 

\sparagraph{Clustering Data.} In order to form abstractions that are \emph{shared} between examples, the refactoring LLM requires a multi-instance scope, i.e., it must receive a batch of related $(q,p)$ tuples at a time.
We implement this by clustering examples using an embedding of the query $q$. 
Specifically, we embed each query using OpenAI's Ada embedding model~\citep{openai2022ada} and hierarchically cluster the embeddings using  Ward's clustering algorithm \citep{ward.j.1963}, implemented via Scikit-Learn \citep{scikit-learn}. 
This gives us a tree of related examples, which we topologically sort and group into batches of $k$, where $k$ is a hyperparameter (see \cref{append:hyperparams} for all hyperparameter values). 

\sparagraph{Curriculum.} Intuitively, shorter and easier programs should contain abstractions that can be reused in harder, more compositional programs, so we sort examples into a curriculum~\citep{bengio2009curriculum}. 
To approximate difficulty, we sort the batches based on the average length (in tokens) of their queries. 
See \cref{append:preproc} for preprocessing details. 

\subsection{Training} \label{sec:method_train}
\method{}'s training data consists pairs of queries $q$ and primitive programs $p$. 
The training phase outputs: 1) the \emph{Code Bank} ($C$): the library of helper functions abstracted out during training and 2) the \emph{Demo Bank} ($D$): examples of the functions being used.
 As shown in \cref{fig:method}, the training phase is an iterative process where the LLM receives as input a batch of queries and primitive programs and then proposes helper functions that can be abstracted (Stage 1). 
For each batch, candidate helper functions are evaluated based on the correctness of the refactored programs that occur in (Stage 2a).
After verification, failed examples are isolated into a second batch and re-attempted (Stage 2b).
To improve the quality of the Code Bank, we periodically edit helper functions after a fixed number of batches to improve their pass rate (over unit tests) and prune ineffective helpers (Stage 3). 
After training, the library of helper functions (Code Bank $C$) is stored for use during testing 
along with successful demonstrations of programs using helper functions (Demo Bank $D$).
Note that the training process can be repeated over multiple epochs. 
Below we describe each stage in detail, with the overall algorithm detailed in \cref{alg:train}.

\textbf{Stage (1): Refactoring Examples (\fname{refactorBatch}).} The main module of the refactoring process takes as input a batch of examples, a set of instructions, and the current set of helper functions in the code bank (if any). 
It prompts the refactoring LLM for a set of new helper functions $H^{new}$ along with refactored versions of each program that uses helpers from $H^{new}$ when appropriate. 

\sparagraph{Stage (2a): Verification (\fname{verify}).} To avoid introducing errors, we need to verify the helper functions and refactored programs generated by the LLM by executing them and comparing the results to the original, i.e., determining if $\hat{p}\text{()} = p\text{()}$. 
The refactored program $(q, \hat{p})$ is stored as a demonstration 
for future use by the agent (cf. \cref{sec:method_test}) if it passes verification. 
Only helper functions that pass verification are added to $C$.
We also store a record of programs that \emph{failed} verification, as these will be crucial in  \sfname{editCodeBank\text{()}} and \sfname{pruneCodeBank\text{()}}, which improve existing functions and prune functions leading to failures.

\sparagraph{Stage (2b): Feedback-based Retrial (\fname{retry}).} 
If a program fails to pass verification, we optionally retry the refactoring process.
In a follow-up prompt, we present failed programs and their helper functions. We also include environment feedback from execution (i.e., the output or error message produced).\footnote{We do not include the output for LOGO as it is an image.} 
The refactoring LLM then produces a new version of each failed program; these are verified and their helpers are added to $C$ if correct. 

\sparagraph{Stage (3a): Editing Code Bank (\fname{editCodeBank}).} 
From the \sfname{verify\text{()}} module, some helper functions fail to pass all unit tests because they contain incorrect abstractions. For example, a function like \sfname{draw\_triangle\text{()}} might start with a hardcoded value for a small size, leading it to fail on medium triangles. 
To update such functions, we construct a prompt for each function in $D$ that shows the LLM passing and failing unit tests and asks it to propose edits to the function; this occurs once every \sfname{editEvery} iterations, where \sfname{editEvery} is a hyperparameter. 
We replace a function if it passes more unit tests after editing.

\sparagraph{Stage (3b): Pruning Code Bank~(\fname{pruneCodeBank}).} 
In this module, we prune helper functions added to $C$ that fail a majority of unit tests and cannot be improved further via editing. 
For each function, we derive a score based on the success rate of programs using the function; we set a threshold below which functions are pruned (shared by all domains).
See \cref{append:codebank_filter} for further details.

We use the dev set to select hyperparameter values, reported in \cref{append:hyperparams}.
All prompts can be found in \cref{append:prompts}.

\vspace{-0.5em}
\subsection{Testing} \label{sec:method_test}
\vspace{-0.5em}
At test time, we deploy a program synthesis -- or semantic parsing -- \emph{agent} that makes predictions for test examples, one at a time.
Unlike related work on using learned tools \citep{yuan.l.2023craft, qian2023creator, wong.l.2023learning}, we explore a variety of open-source LLMs, in addition to a black-box LLM (GPT-3.5). 
Following effective strategies in semantic parsing and in-context learning~\citep{shin.r.2022, roy.s.2022, bogin.b.2023, liu-etal-2022-makes, yasunaga2023large}, for each test example, the agent constructs a prompt with in-context learning (ICL) examples retrieved from a training corpus, followed by a test query.
The examples are retrieved from the training data using vector similarity between the training queries and the test query.
Further details in \cref{append:testing}.

\looseness -1

\textbf{\method{}-augmented Agent.}  Our agent has access to the \emph{training data} and \emph{code bank}, as well as examples of refactored programs in the \emph{demo bank}.
The ICL budget (10 examples for all experiments) is split between primitive training examples and refactored ones.\footnote{We found it necessary to keep some primitive programs as ICL examples, as not all test queries can be handled by helper functions alone. We treat the ratio of primitive to refactored programs in the ICL example as a hyperparameter (all values listed in \cref{tab:hyperparams}).}
In addition to these demonstrations, the augmented agent retrieves up to 20 relevant helper functions from the code bank, where relevance is measured by the similarity between the query and the function name and description. 
These helper functions are concatenated into the prompt. 
The final input is a prompt containing the instructions, the retrieved helper functions, the mixed ICL examples, and the test query. 
To encourage the model to use helper functions, we include a ReAct-style prompt \citep{yao2023react} that first asks the model to \emph{think} about which functions might be relevant based on the query and then generate the code.\footnote{Without these additional ``thought'' statements, we found the augmented agent rarely uses any helper functions.} 

\looseness -1
\vspace{-0.5em}
\section{Experimental Setup} \label{sec:expts}
\vspace{-0.5em}
\looseness -1
\subsection{Datasets}
We explore five datasets: LOGO, Date understanding, TextCraft, MATH, and TabMWP.
A common thread through these datasets is that they contain heterogenous problems requiring multiple helper functions as opposed to problems like sorting, which are challenging for LLMs but can be solved with a single function~\citep{dziri2023faith}. 
Statistics for the datasets are given in \cref{tab:data_stats}. 
See \cref{append:data} for details about each dataset and its primitive operations.

\sparagraph{LOGO.} LOGO is based on the Logo Turtle graphics domain-specific language \citep{abelson.h.1986turtle}, with which basic graphics can be drawn by controlling a pen (the ``turtle'') that draws as it moves through space, using commands like \sfname{forward(dist)} and \sfname{left(theta)}. 
The data we use is based on \citet{ellis.k.2021}'s LOGO dataset, re-annotated by \citet{wong.l.2021}. 
For easier prediction by LLMs, we parse the data into abstract syntax trees and write a set of rules for translating these into Python; we release this rewritten data.
We use the ``small'' train/test splits (200/111) from \citet{wong.l.2021} and take $100$ dev examples from the ``large'' train set. 

\sparagraph{Date.} 
We use the date understanding task from BigBench-Hard \citep{srivastava.a.2022bigbench, suzgun.m.2022challenging}, which consists of short word problems requiring date understanding.
We obtain silver programs from \citet{lyu.q.2023faithful}'s predictions. 
Specifically, we split their predicted programs from GPT-3.5 into train, dev, and test splits (66/113/180) and filter the train split by correctness.

\begin{table*}[ht]
\centering
\setlength{\tabcolsep}{2pt}

    \caption{Accuracy of baseline agents predicting primitive programs (Prim.) and those augmented with \method{} helper functions (3 random seeds). 
Across domains and models, \method{} improves over a strong baseline agent with access to the same number of ICL examples. 
Math domains with no clear domain language marked with $^*$.}
    \vspace{0.1in}
\begin{tabular}{lcc|cc|cc|cc|cc}
\toprule
 &  \multicolumn{2}{c}{\bf LOGO} &  \multicolumn{2}{c}{\bf Date} &  \multicolumn{2}{c}{\bf TextCraft} &  \multicolumn{2}{c}{\bf MATH (Alg.)$^*$} &  \multicolumn{2}{c}{\bf TabMWP$^*$} \\
\cline{2-11} \cline{4-5} \cline{6-7} \cline{8-9} \cline{10-11} \\  [-1.5ex]
\textbf{Agent} & \bf Prim. & \bf \method{} & \bf Prim. &\bf  \method{} & \bf Prim. &\bf \method{} & \bf Prim. & \bf \method{} & \bf Prim. & \bf \method{} \\
\midrule
 CodeLlama-7B & $34.5\interval{1.3}$ & $34.5\interval{1.6}$ & $52.4\interval{0.7}$ & $55.2\interval{1.4}$ & $12.8\interval{1.3}$ & $16.7\interval{1.3}$ & $6.8\interval{0.8}$ & $17.6\interval{0.8}$ & $16.7 \interval{1.6}$ & $27.7\interval{0.6}$ \\
 CodeLlama-13B & $45.6\interval{0.3}$ & $\mathbf{57.1\interval{0.6}}$ & $42.8\interval{2.0}$ & $68.9\interval{1.6}$ & $18.8\interval{0.7}$ & $26.9\interval{2.2}$ & $14.0 \interval{0.9}$ & $17.6\interval{0.8}$ & $29.1\interval{0.6}$ & $27.9\interval{0.4}$\\
 CodeLlama-34B & $50.2\interval{0.8}$ & $50.8\interval{0.6}$&  $47.2\interval{1.5}$ & $68.5\interval{2.1}$ & $22.2\interval{0.7}$ & $\mathbf{30.8\interval{1.3}}$ & $14.9\interval{1.1}$ & $22.5\interval{1.6}$ & $19.4\interval{1.2}$ & $27.0\interval{0.8}$ \\
\midrule
 Lemur-70B & $44.1\interval{1.4}$ & $56.8\interval{0.9}$ & $68.2\interval{0.4}$ & $70.5\interval{0.6}$  & $15.7\interval{1.7}$ & $23.5\interval{2.1}$ & $14.0\interval{1.2}$ & $13.1\interval{1.2}$ & $27.5\interval{0.9}$ & $25.7\interval{0.8}$ \\ 
\midrule
 GPT-3.5-turbo & $36.9\interval{1.6}$ & $49.3\interval{1.1}$ & $88.9\interval{0.3}$ & $\mathbf{90.2\interval{0.5}}$  & $15.4\interval{1.3}$ & $18.4\interval{2.0}$ & $43.2\interval{1.6}$ & $\mathbf{55.4}\interval{2.8}$ & $87.4 \interval{0.9}$ & $\mathbf{89.2}\interval{0.8}$ \\
\bottomrule
    \end{tabular}
    \label{tab:main_res}
\end{table*}

\sparagraph{TextCraft.} To explore the utility of \method{} in LLMs-as-agent settings~\citep{liu2023agentbench}, we use the TextCraft dataset~\citep{prasad2023adapt} that requires an agent to craft Minecraft items within a text-only environment~\citep{cote2019textworld}. Each task instance in TextCraft consists of a goal (query) and a series of 10 crafting commands that contain recipes for related items including distractors.  
Unlike \citet{prasad2023adapt}, who use TextCraft in an interactive setting where the LLM agent receives textual feedback from the environment at each step, 
we ask the agent to generate a single Python program for executing the entire task at once, making the task \emph{more challenging}. 
We evaluate on the depth 2 split of the test set used in \citet{prasad2023adapt} while using a subset of depth 1 recipe examples for our dev set, giving us a train/dev/test split of 190/50/77.

\sparagraph{MATH.} To test \method{}'s ability to discover functions in domains that do \emph{not} have a pre-defined domain language, we additionally examine the MATH dataset \citep{hendrycks2021measuring} consisting of challenging math word problems. 
As with Date, we generate silver training programs; in this case, we use GPT-4 with a Program-of-Thoughts prompt \citep{chen.w.2022program}.
To have a good chance of generating sufficient numbers of training programs, we use examples from hardness levels 1 and 2 (MATH contains 5 levels of difficulty).
We also focus on the Algebra subset of MATH, which is the largest.
This gives us a train/dev/test split of 194/61/74.
While only correct programs are used for training, correct and incorrect examples are used in dev and test. 
Note that unlike the other datasets, MATH does not have an inherent domain language associated with it and does not have any primitives beyond the ones already contained in Python.

\sparagraph{TabMWP.} We further extend our general experiments on MATH by testing on TabMWP \citep{lu2022dynamic}, a tabular resoning dataset consisting of math word problems about tabular data. 
Here, we also use silver programs for training and focus on levels 1-4 (TabMWP has 8 difficulty levels), following a similar procedure as for MATH. This gives us a train/dev/test split of 194/60/74.

\subsection{Baselines}
\label{ssec:baselines}
\sparagraph{Baselines from Prior Work.} 
We compare \method{} against relevant external baselines from past work. However, note that multiple methodological differences in our work, like the use of ICL examples and the format of the output programming language, give our agent an inherent advantage over these baselines. 
Thus, we refer to these numbers primarily to highlight the strength of our baseline agent.
For LOGO, we use the ``offline synthesis'' numbers reported by \citet{grand.g.2023}, which resemble our train/test setting; however, we note that \citet{grand.g.2023} predict programs in their original Lisp format and use a different agent model.
For the Date dataset, we run \citet{lyu.q.2023faithful}'s Faithful-CoT method on our custom test split using \sfname{gpt\text{-}3.5\text{-}turbo}. 
While the output format and models used are the same, both our baseline and \method{} use retrieved examples for in-context learning, while \citet{lyu.q.2023faithful} do not. Furthermore, our ICL examples are based on programs generated by \citet{lyu.q.2023faithful} after filtering for correctness, leading to better performance even from our baseline agent. 
Finally, for TextCraft we re-run \citet{prasad2023adapt}'s baseline -- based on ReAct \citep{yao2023react} -- on the depth-2 test set of TextCraft. 
Here, we use \sfname{gpt\text{-}3.5\text{-}turbo\text{-}instruct}, as \citet{prasad2023adapt} found it to outperform \sfname{gpt\text{-}3.5\text{-}turbo}. 

\sparagraph{Baseline Programming Agent.}
For a more direct comparison, we implement a baseline agent that has access to all the same data as \method{} but does not use abstractions, thus directly testing the role of \method{} abstractions in performance. 
Our baseline agent retrieves primitive programs from the training data; note that this is exactly \emph{the same dataset} used for refactoring, i.e., the baseline LOGO agent retrieves demonstrations from the LOGO training examples.
The input to the baseline agent is a prompt with the same overall instructions as the \method{} agent (including a description of the primitives), the ICL examples, and the test query; the output is a program for the test query.
We use a fixed budget of 10 ICL examples so that the baseline agent sees exactly as many demonstrations as the \method{} agent.

\section{Results}
\label{sec:results}

\textbf{Comparison to External Baselines.} \cref{tab:comp} shows a comparison of the baseline and \method{} agents the external baselines from prior work.
Note that we do not include MATH and TabMWP here as we use a subset of levels for these datasets.
We first note that \method{} outperforms the baselines in all cases. Furthermore, because of the methodological differences detailed in \cref{sec:expts}, our baseline ``primitive-only'' agent -- equipped with ICL examples and using a code-specific LLM -- also outperforms past baselines on LOGO and Date.
On TextCraft, the ReAct baseline from \citet{prasad2023adapt} has an advantage in that it receives environmental feedback, while our baseline does not. 
Nevertheless, even \emph{without feedback} \method{} outperforms ReAct. 
Thus, we compare primarily against our baseline agent, as this provides a direct measure of the impact of abstractions (rather than the other changes made). 

\begin{table}[ht]
    \centering
    \setlength{\tabcolsep}{2pt}
    \caption{Comparison to relevant past work in each non-math domain. TC$^{\dagger}$ denotes the TextCraft dataset.}
    \vspace{0.1in}
    \begin{tabular}{lllc}
    \toprule
    & \bf Method & \bf Agent & \bf Acc. \\
    \midrule
        \multirow{3}{*}{\rotatebox[origin=c]{90}{ \footnotesize{LOGO}}} &\textsc{Lilo} & Code-davinci & $41.1$\\
        & Primitive Programs & CodeLlama-13B & $45.6$ \\
        & \method{} Programs  & CodeLlama-13B & $\mathbf{57.1}$\\
    \midrule
    \multirow{3}{*}{\rotatebox[origin=c]{90}{ \footnotesize{Date}}} & Faithful-CoT & GPT-3.5-turbo& $83.3$\\
        & Primitive Programs &  GPT-3.5-turbo & $88.9$\\
        & \method{} Programs &  GPT-3.5-turbo & $\mathbf{90.2}$\\
\midrule
\multirow{3}{*}{\rotatebox[origin=c]{90}{ \footnotesize{TC$^{\dagger}$}}} & ReAct &  GPT-3.5-turbo & $25.6$\\
        & Primitive Programs & CodeLlama-34B  & $22.2$\\
        & \method{} Programs &  CodeLlama-34B & $\mathbf{30.8}$\\
    \bottomrule
        
    \end{tabular}
    \label{tab:comp}
    \hfill
\end{table}

\textbf{\method{} outperforms the baseline agent in non-math domains.} \cref{tab:main_res} shows \method{}'s performance compared to the baseline agent using primitive programs (described in \cref{ssec:baselines}).
Overall, for each model type, \method{} generally outperforms the baseline by a wide margin; 
for example, \method{} provides CodeLlama-13B a $11.5\%$ boost on LOGO, allowing it to outperform much larger models. 
Across datasets, CodeLlama-13B generally benefits most from \method{} abstractions. 
\cref{tab:main_res} also shows that large models also benefit from \method{}, with large gains for Lemur-70B and GPT-3.5 on LOGO and TextCraft. 
Finally, the largest models are not necessarily the best: on LOGO and TextCraft, GPT-3.5 is outperformed by open-source models, especially after augmentation, e.g., CodeLlama-13B with \method{} abstractions is able to outperform GPT-3.5 \emph{without} abstractions by $19.2\%$ (it also outperforms GPT-3.5 with abstractions).
Thus, by running \method{}'s training process on only $\sim\!200$ examples, we can bring a much smaller open-source model's accuracy far beyond that of a (likely) much larger system. 

\textbf{\method{} improves over the baseline on mathematical domains.}
Our math domains (MATH and TabMWP) differ from the others in that there is no clear domain language that \method{} should discover.
Unlike a domain like LOGO (where it is clear which functions a human programmer would write), there is less clarity on which helper functions should be discovered. 
Nevertheless, in this challenging setting \method{} generally improves over the baseline agent across different models, with major gains for GPT-3.5 on MATH ($11.8\%$) and for CodeLlama-7B on TabMWP ($11\%$), where the 7B \method{} performance is comparable to the 13B performance.
Overall, Lemur struggles with these problems, producing a large number of empty outputs both for the baseline and \method{} agents; when taking the standard error into account, baseline and \method{} performance is roughly comparable.
These results highlight \method{}'s flexibility and ability to generalize to domains without clear-cut domain languages.

\textbf{Ablations.}
In \cref{sec:method_train} we describe \method{}'s multiple components; here we determine the utility of each by removing each one in isolation for non-math datasets.
\cref{tab:ablations} shows the results of these ablations.
We use the CodeLlama-13B agent due to the size of \method{}'s impact on it across tasks. 
We average the performance across 3 seeds. 
\cref{tab:ablations} shows that each component contributes to performance, with drops when any is removed. 
Across datasets, the largest performance decreases come with the removal of retrials and with removal of the curriculum.
Retrials can not only increase the number of useful helper functions but can also help increase the number of examples in the Demo Bank. 
Replacing the curriculum with a random ordering also severely hurts performance, e.g., leading to an $18.7\%$ drop on LOGO. 

\begin{table}[t]
    \small
    \centering
    \caption{Ablations of each optional \method{} component tested on dev splits with CodeLlama-13B.
    To remove the curriculum, we randomly shuffle example clusters instead of presenting them in order of shortest query to longest query. }
    \vspace{0.1in}
    \begin{tabular}{lccc}
    \toprule
    \textbf{Ablation} & \textbf{LOGO} & \textbf{Date} & \textbf{TextCraft} \\
    \midrule 
    \method{} & 55.0 & $77.0$ & $34.12$\\
    \midrule
    \quad -- \sfname{retry} & $48.3$ & $51.9$ & $30.43$ \\ 
    \quad -- curriculum &  $36.3$ & $56.6$ & $28.78$\\
    \quad -- \sfname{pruneCodeBank} & $52.0$ & $65.5$ & $25.26$ \\
    \quad -- \sfname{editCodeBank} & $53.3$ & $69.6$ &  $27.35$\\
    
    \bottomrule
    \end{tabular}
    \label{tab:ablations}
\end{table}

\textbf{\method{} learns general, reusable functions.}
In \cref{sec:intro}, we stressed the importance of reusability. 
Specifically, generating programs without shared abstractions means that the model has to re-generate subprograms that could be reused across multiple test instances.
We argue that \method{} improves over this paradigm by learning \emph{shared} abstractions.
The results in \cref{tab:main_res} indicate that \method{} offers large improvements over a baseline agent that lacks abstractions. Here, we verify that the abstractions learned are reusable, i.e., shared. 
\cref{fig:usage} shows the number of times the top-5 most common \method{} functions are called in test programs produced by the CodeLlama-13B agent. 
Across all datasets, we see that the helper functions learned by \method{} are commonly reused, with the most relative reuse in TextCraft.
\cref{append:qual} shows examples of these common functions.

\section{Analysis} \label{sec:analysis}

\subsection{What kinds of programs are discovered?}
\begin{figure}[t]
    \centering
    \includegraphics[width=0.9\linewidth]{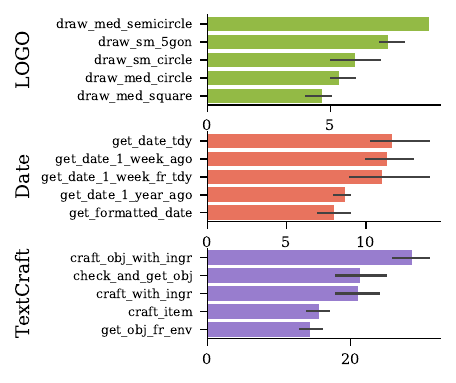}
    \vspace{-1em}
    \caption{Function usage by CodeLlama-13B for the top-5 most common helpers illustrating reusability across examples. The x-axis denotes the number of times a functions is used in the test set.}
    \label{fig:usage}
\end{figure}

To further examine what kinds of helper functions are discovered, we examine the most frequent helper functions for each domain from \cref{fig:usage}, summarizing the results below. Refer to \cref{append:qual} for the implementation of these functions. We find that distinct trends emerge across domains. 

For LOGO, \method{} discovers functions that encapsulate different \emph{types of shapes}. 
This is expected, as the LOGO data was generated with these functionalities in mind, i.e., the larger shapes are composed of objects like semicircles, pentagons, and circles. 
For Date, \method{} tends to encapsulate single operations, prioritizing \emph{interpretablity in function names} like \sfname{get\_date\_one\_year\_ago\text{()}}. While seemingly less complex than LOGO's functions, this approach aligns with the importance of function naming in synthesis procedures, as highlighted by \citet{grand.g.2023}. 
In TextCraft, the functions uncovered by \method{} are more \emph{complex} and reflect the \emph{dynamics of the game}. 
Specifically, the functions include conditional statements for checking ingredients, reflecting the fact that in TextCraft, having the correct crafting ingredients is a prerequisite for crafting an object (see \sfname{craft\_and\_get\_ingredient\text{()}} in \cref{fig:method} and \cref{tab:tc_programs}, which is taken from the learned code bank $C$). 
\subsection{What kinds of errors do agents make?} \label{sec:error}
To better understand how \method{} aids program generation and also examine cases where it does not help, we perform a two-part error analysis. 
First, we examine cases where the \method{}-augmented program was correct and the baseline agent's primitive program was incorrect. 
We then examine the opposite set of cases, where the baseline program was correct but the \method{} program was incorrect. 

\cref{fig:fig1} shows the first kind of comparison on LOGO using the CodeLlama-13B model, where we qualitatively show an actual example that highlights the benefits of reuse and abstraction. 
The baseline program makes an error in calculating the polygon's interior angle when generating the program from scratch.
This is avoided by the \method{} agent, which simply uses a verified helper to generate the polygon correctly. 
The example also illustrates the importance of abstraction: as queries become more complex, generating a solution from scratch becomes more challenging. 
The baseline program reasons incorrectly about code outside of the shape, failing to use \sfname{embed\text{()}} correctly.
Meanwhile, the augmented program offloads reasoning about the shape to an easily-matched function, and is able to correctly use \sfname{embed\text{()}}. 
To quantify these trends, we manually inspect the output of the baseline CodeLlama-13B on LOGO on the 25 cases where the \method{} agent was correct, categorizing them into errors involving reasoning (first example in \cref{fig:fig1}) and shape-internal errors (second example); we find 16 reasoning and 9 shape errors. 
We also examine \method{}'s failure modes by manually inspecting all cases where the augmented agent failed and the baseline succeeded, again using CodeLlama-13B on LOGO; there are 13 such cases. 
We categorize them into three types: 
\begin{itemize}[noitemsep,nolistsep,topsep=0pt,leftmargin=*]
    \item \textbf{Incorrect connector code}: (7 exs.) the program fails due to mistakes in the primitive operations or control flow.  
    \item \textbf{Incorrect/undefined function}: (4 exs.) the code refers to non-existent functions, or incorrectly calls a function similar to the correct one.
    \item \textbf{Verification failure}: (2 exs.) the program was correct but the verification function gives a  false negative.\footnote{For example, for \emph{``a small square next to a small 6 gon''} the agent generates the hexagon to the left of the square, where in the reference it is to the right.} 
\end{itemize}
Thus, the most common error is a failure to predict primitives; here, the \method{} agent is at a disadvantage w.r.t. the baseline, as they share the same ICL budget. 
The baseline agent sees 10 examples with \emph{only} primitive code, while the \method{} agent sees 5 primitive and 5 Demo Bank examples.

\begin{figure}
    \centering
    \includegraphics[trim={0.2cm 0.1cm 0.1cm 0.1cm}, clip, scale=0.95]{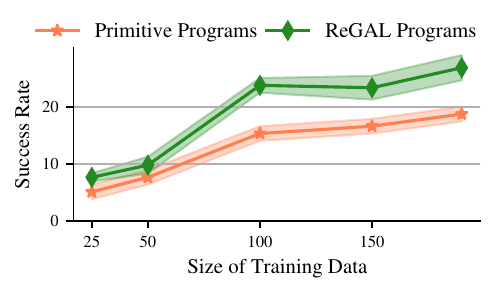}
     \vspace{-1em}
    \caption{
    \method{} programs yield a higher success rate (accuracy) compared to primitive programs on TextCraft for different sizes of training set $X$ using CodeLlama-13B. }
    \label{fig:sample}
\end{figure}

\subsection{How many training examples does \method{} need?} \label{sec:efficiency}
As mentioned in \cref{ssec:baselines}, both baseline and \method{} agents rely on demonstrations of queries and gold programs ($X$) to retrieve most similar ICL examples. 
Additionally, \method{} uses the same demonstrations to learn helper functions in the code bank $C$. We now study how the performance of both agents scales with the size of annotated gold programs in train set $X$ using the CodeLlama-13B model on TextCraft. 
From \cref{fig:sample}, we observe that the \method{} agent consistently outperforms the baseline agent as we vary the number of training examples. Notably, helper functions learned by \method{} yield a $2.56\%$ improvement with as few as 25 demonstrations and an $8.45\%$ improvement with nearly half the size of the train set used in \cref{sec:results}. Additionally, the performance of both the baseline and \method{} agents improves as the number of demonstrations increases. This is expected as both agents benefit from the retrieval of demonstrations similar to the test query as the training set becomes larger and consequently more diverse.

\subsection{How can \method{} adapt to distribution shifts?}
\label{sec:adapt}
To assess the usefulness of abstractions learned by \method{} under a distribution shift at inference-time, we use the TextCraft dataset. We redistribute the instances into three splits: train (seen), dev (unseen), test (unseen) consisting of 100/20/76 instances, such that the dev and test sets contain objects \emph{never encountered} during training. For example, if the test set requires crafting ``yellow jungle fence'', the train set does not contain any instance with a ``fence'' as a target object. Furthermore, we study how to adapt the learned library with unseen examples from the dev set using \method{} in two ways: (i) only pruning the learned library to retain generalizable functions (Stage 3(b) of \method{}); and (ii) an additional iteration of \method{} training.

\begin{table}[t]
    \small
    \centering
    \setlength{\tabcolsep}{2pt}

    \caption{Performance of baseline and \method{} agent under distribution shift on TextCraft with unseen objects in the test set.}
    \vspace{0.1in}
    \begin{tabular}{lcc}
    \toprule
    \textbf{Method} & \textbf{Adaptation Setting } & \textbf{Acc.}\\
    \midrule 
    Prim. Programs & $-$ & 7.8\\
    \method{} & $-$ &  11.4\\
    \method{} & \sfname{pruneCodeBank} on dev (unseen) & 14.9\\
    \method{} & All \method{} stages on dev (unseen)& 17.2\\
    \bottomrule
    \end{tabular}
    \label{tab:adapt}
\end{table}

The results are presented in \cref{tab:adapt}. First, we observe that the performance of both baseline and \method{} agents degrades under distribution shift because the in-context examples may not be adequately relevant to the test instance. However, ReGAL outperforms the baseline agent with primitive programs by $3.6\%$. Furthermore, simply pruning an existing code bank increases performance of ReGAL agent by an additional $3.5\%$. Note that this still uses a learned codebank from prior training and without proposing any new abstractions. 
Finally, learning new abstractions and editing existing ones based on unseen dev data is the most effective yielding an additional $2.3\%$  over only performing \sfname{prunceCodeBank} on the unseen data for a total of $9.4\%$ improvement over the baseline.

\vspace{-0.5em}
\section{Discussion}
\vspace{-0.5em}
\sparagraph{Fixed vs. Variable Costs.} In \cref{sec:results}, \method{} was especially effective for open-source LLMs like CodeLlama.
This result is encouraging, as it indicates that we can bring freely available and open-source models up to at least the same performance as a proprietary, closed-source model (if not more) using \method{} abstractions. 
Thus, we can convert a variable cost -- running an LLM on test data, which scales linearly with the size of the test set -- into the fixed cost of running \method{} to learn a library of helper functions. 

\sparagraph{Connections to Semantic Parsing.}  Executable semantic parsing \citep{winograd.t.1972, zelle.j.1996} typically involves mapping queries to a domain-specific language (DSL), a set of abstractions for a particular application, e.g., SQL operations for querying databases.
These DSLs are manually defined; one way to view \method{} is as a way of \emph{learning} a DSL on top of an extremely general set of primitives. 
A key benefit of \method{} is its generality: on five different domains, it learns useful abstractions without human intervention, while, in a standard semantic parsing setting, these abstractions would designed by hand.

\sparagraph{Connections to Hierarchical Reinforcement Learning.} 
Another way to view the functions discovered by \method{} is as low-level policies composed of primitive actions.
In this view, \method{} resembles hierarchical reinforcement learning \citep[HRL;][]{barto.a.2003hrl}, where tasks are split into skill selection by a controller and the skill policies themselves. 
Our agent LLM acts as a controller while \method{}'s training stage is responsible for discovering a useful set of skills; this is akin to option discovery \citep{sutton.r.1999option}.
While \method{} has a similar hierarchy, it differs in that its skills are symbolic, interpretable, and editable, as opposed to HRL policies, which typically are not.

\textbf{Limitations. }
As mentioned in connection to HRL, the functions \method{} learns are code-based. 
This can make them less flexible than functions parameterized by neural networks \citep[e.g.,][]{andreas.j.2016}, especially in domains where the environment can change dynamically, e.g., navigation tasks.  
However, \method{}'s verification-based pruning means that no functions would be discovered in these cases. 
Relatedly, not every domain has reusable abstractions, and not every example stands to benefit from them; the primitives for a domain may already be suitably abstract, e.g., if they already form a DSL. 
Finally, in \cref{append:qual} we see that \method{}'s abstractions are not necessarily the same as those a human would choose.

\vspace{-1em}
\section{Conclusion.} 
We introduce \method{}, a gradient-free approach to learning abstractions from a small set of examples. 
Our experimental results show that abstractions from \method{} improve the accuracy of programs predicted by a variety of LLMs across five diverse domains. 
Furthermore, \method{} abstractions are reusable and general, allowing them to be applied across examples for a given task. 
In our analysis, we find that the functions learned by \method{} codify commonly-used subroutines as well as task dynamics.
Our error analysis indicates that \method{}'s improvements come both from function reuse as well as simplification of the reasoning involved in program prediction.

\section*{Acknowledgements}
We thank Yichen Jiang, Justin Chen, Jaehong Yoon, and Swarnadeep Saha, as well as the anonymous reviewers, for their valuable feedback on the paper. This work was supported by NSF-AI Engage Institute DRL-2112635, DARPA Machine Commonsense (MCS) Grant N66001-19-2-4031, and the Accelerate Foundation Models Research
program. The views contained in this article are those of the authors and not of the funding agencies.

\section*{Impact Statement}
Our work aims to learn symbolic functions given a set of demonstrations; this has the potential to improve LLM predictions not only in terms of accuracy but also in terms of interpretability and trustworthiness.
Unlike the mechanisms of an LLM itself, a Python function is natively interpretable by a human and can be debugged. Furthermore, results obtained by executing such a function are inherently faithful, in that we can identify the exact trace of operations that generated the result~\citep{lyu.q.2023faithful}. 
Our work does not have more potential for negative use than typical LLM-based systems and is subject to the biases inherent to these models and the datasets they are trained on~\citep{weidinger2021ethical}. 
As with any system generating code, particular caution should be taken before executing snippets with the potential to damage the execution environment~\citep{ruan2023identifying}.

\bibliography{refactoring}
\bibliographystyle{icml2024}

\appendix
\section*{Appendix}
\section{Methods}
\subsection{Preprocessing} \label{append:preproc}
\textbf{Adding comments.} \label{append:comments}
To add comments, we first use a zero-shot prompt to break the query down into its constituent parts; for example, a LOGO query like \emph{``Place 4 small semicircles in a row''} is broken down into \emph{``1. place semicircles 2. small semicircles 3. in a row 4. 4 semicircles}. 
We then include this decomposition in a prompt asking the model to add comments to the code. 
After adding comments, we verify the code first with exact match (excluding comment strings) and then use execution accuracy if exact match fails. 

\subsection{Training} \label{append:training}
\paragraph{Code Bank Editing.} \label{append:codebank_refactor}
Our Code Bank editing prompt asks the model to produce a CoT-style output, first specifying \emph{why} the failing unit tests failed and then proposing an edit for the function. 
We then execute the stored demonstrations for that function with the new version; if there are more passing cases after refactoring, we replace the function. 
If the new function's signature differs from the old, we use a simple prompt to refactor the unit tests to accommodate the new function. 

\begin{table}[ht]
    \small
    \centering
    \caption{Dataset statistics. We list the number of primitive operations in the programs (aside from built-in Python functions).}
    \vspace{0.1in}
    \begin{tabular}{l|cccc}
    \toprule
    \textbf{Dataset} & \textbf{Train} & \textbf{Dev} & \textbf{Test} & \textbf{\# Primitives} \\
    \midrule
    LOGO & 200 & 100 & 111 & 9  \\
    Date & 66 & 113 & 180 & 4\\
    TextCraft & 190 & 50 & 77 & 3 \\
    MATH (Alg.) & 194 & 61 & 74 & 0 \\
    TabMWP & 194 & 60 & 74 & 0 \\
    \bottomrule
    \end{tabular}
    \label{tab:data_stats}
\end{table}

\paragraph{Code Bank Pruning.} \label{append:codebank_filter}
For each function, given a set of passing programs $P$ and failing programs $F$, we compute a score $s = |P|  - \sum_{p \in F} 1/n_{p}$, where $n_{p}$ is the number of functions used in $p$. 
In other words, the function receives $+1$ for each passing program it participates in, and a negative score inversely proportional to the number of functions in the program (since naively, the failure could be attributed to any of the functions). 
Functions are pruned if they have been used a sufficient number of times and $s$ falls below a threshold $\theta$ (set to 0 for all experiments). 

\subsection{Testing} \label{append:testing}
In our test-time agent, we use ChromaDB\footnote{\url{https://github.com/chroma-core/chroma/}} for indexing and retrieval with OpenAI Ada embeddings. 
ICL examples are retrieved from the training data and from the Demo Bank using query similarity.
We limit the number of Code Bank functions to 20, using the similarity between the query and the function name for retrieval. 
The Code Bank is pruned once before testing. 

\subsection{Models} \label{append:models}
For GPT-3.5, we use the \sfname{gpt\text{-}3.5\text{-}turbo} version (0613). 
All CodeLlama models use the \sfname{CodeLlama\text{-}*\text{-}Instruct\text{-}hf} versions, and we use the \sfname{lemur\text{-}70b\text{-}v1} version of Lemur. For the latter two open-source models, we use the checkpoints from HuggingFace~\citep{wolf2019huggingface}.

\begin{algorithm}[ht]
\caption{\method{}: Training Algorithm}
\label{alg:train}
\begin{algorithmic}{}
    \State {\bfseries Input:} $X = (q, p)$ \textcolor{comm}{\footnotesize{\textit{// Training data: (query, program)}}}
    \State {\bfseries Params:}  \textrm{editEvery}, \textrm{pruneEvery}, threshold $\theta$
    \State {\bfseries Output:} CodeBank $C$, DemoBank $D$
    \State $C \gets \varnothing, D \gets \varnothing$  \textcolor{comm}{\footnotesize{\textit{// Initialization, i.e., no refactoring}}}
    \State \textcolor{comm}{\footnotesize{\textit{// Preprocessing data via clustering and sorting by difficulty}}} 
    \State $\mathcal{P} \leftarrow \boldsymbol{\tt{preprocessAndGroupData}}(X)$ 
    \For{ index $g$, batch $\mathcal{G} \in \mathcal{P}$}
        \State \textcolor{comm}{\footnotesize{\textit{// Refactor programs in group $\mathcal{G}$ based on the current CodeBank $C$. Returns new programs and helper functions.}}} 
        \State $(p^{new}_1, h_1),\!\ldots,\!(p^{new}_k, h_k) \!=\! \boldsymbol{\tt{refactorBatch}}(\mathcal{G}, C)$ 
        \State $H^{new} \gets \{ h_1, \cdots, h_k \}$  \textcolor{comm}{\footnotesize{\textit{// Set of new helper functions $H$}}}
        \State \textcolor{comm}{\footnotesize{\textit{// Verifying that the gold program and the refactored program yield the same result when executed via indicator $\delta^{new}.$}}} 
        
        \State $\delta^{new}_{1:k} \leftarrow \boldsymbol{\tt{verify}}(H, C, \{p^{new}_i\}_{i=1}^k, \{p_i\}_{i=1}^k)$ 
            \For{ $i \in \{i: \delta^{new}_i = False\}$}
                \State $(p^{retry}_i, h^{retry}_i) \leftarrow \boldsymbol{\tt{retry}}(p_i, p^{new}_i, C)$ 
                \State $\delta^{new}_i \leftarrow \boldsymbol{\tt{verify}}(h^{retry}_i \cup H, C, p^{new}, p)$
                \If{$\delta^{new}_i\!\!=\!\!True$}
                \textcolor{comm}{\footnotesize{\textit{// Update if retry succeeds}}}
                \State $p^{new}_i \leftarrow p^{retry}_i $ 
                \State $ H^{new}[i] \gets h^{retry}_i$
                \EndIf
            \EndFor 
        \State \textcolor{comm}{\footnotesize{\textit{// update CodeBank $C$ with successful helper functions}}}
        \State $C \leftarrow C + H^{new}[i]$ \textbf{for} $i \in \{i: \delta^{new}_i = True\}$
        \State \textcolor{comm}{\footnotesize{\textit{// update DemoBank $D$ with refactored programs}}}
        \For{ $i \in \{1,\ldots, k\}$} 
            \State$D \leftarrow D  + (p^{new}_i, \delta^{new}_i)$ 
        \EndFor
        \State \textcolor{comm}{\footnotesize{\textit{// edit and prune CodeBank }}}
        \If{ $g \pmod{\mathrm{editEvery}} = 0$} 
            \State $C \leftarrow \boldsymbol{\tt{editCodeBank}}(C, D)$
        \EndIf
        \If{ $g \pmod{\mathrm{pruneEvery}} = 0$} 
            \State $C, D \leftarrow  \boldsymbol{\tt{pruneCodeBank}}(C, D, \theta)$
        \EndIf
    \EndFor
    \Return{$C,D$}
\end{algorithmic}
\end{algorithm}

\begin{algorithm}[ht]
\caption{\method{}: Testing Algorithm}
\label{alg:test}
\begin{algorithmic}{}
    \State {\bfseries Input:} $Q$, $C$, $D$, $X$ \textcolor{comm}{\footnotesize{\textit{// Test queries $Q$, Code Bank $C$, Demo Bank $D$, Training data $X = $ (query, program)}}}
    \State {\bfseries Params:}  \textrm{ICL Budget} $M$, \textrm{ICL Percentage} $r$ 
    \State {\bfseries Output:} Predicted programs $\hat{P}$
     
    \State $M^{demo} \leftarrow r * M$
    \State $M^{train} \leftarrow M - M_{demo}$ 
    \State $\hat{P} \leftarrow \varnothing$
    \For{ $q \in X$}
        \State $H \leftarrow \boldsymbol{\tt{retrieve}}(q, C, 20)$ \textcolor{comm}{\footnotesize{\textit{// retrieve up to 20 helper functions conditioned on the query}}}
        \State $X^{demo} \leftarrow  \boldsymbol{\tt{retrieve}}(q, D, M^{demo})$ \textcolor{comm}{\footnotesize{\textit{// retrieve helper demos from $D$}}}
        \State $X^{train} \leftarrow  \boldsymbol{\tt{retrieve}}(q, X, M^{train})$ \textcolor{comm}{\footnotesize{\textit{// retrieve primitive demos from $X$}}}
        \State $I \leftarrow \boldsymbol{\tt{createPrompt}}(H, X^{demo}, X^{train})$
        \State $\hat{p} \leftarrow LLM(I)$ \textcolor{comm}{\footnotesize{\textit{// generate program}}}
        \State $\hat{P} \leftarrow \hat{P} + \hat{p}$
    \EndFor
    \Return{$\hat{P}$}
\end{algorithmic}
\end{algorithm}

\subsection{Data} \label{append:data}

\subsubsection{LOGO} \label{append:logo}

LOGO data is generated from a synchronous text-code grammar, and pairs procedurally-generated language commands like \emph{``three small triangles in a row''} with a corresponding Turtle graphics program; however, the original LOGO dataset is expressed in a Lisp-style functional syntax. 
While this facilitates the application of helpful data structures for efficient code search \citep{ellis.k.2021, bowers.m.2023top}, object-oriented languages like Python are far more common in practice. 
As a result, they are represented more in LLM pretraining data, which has been shown to contribute to parsing performance \citep{bogin.b.2023}. 
To account for this, we write an AST-based parsing script to translate the LOGO dataset into Python. 
The LOGO dataset was released under an MIT License.

\sparagraph{Primitives.}
\cref{tab:logo_primitives} describes the primitives available in the LOGO library. 
Note that these are in addition to all Python primitives. 
We also provide all agents with several hard-coded values for long loops and small steps so that they can draw round shapes. \sfname{HALF\_INF} is the number of steps required to draw a semicircle. \sfname{EPS\_DIST} is a small distance, and \sfname{EPS\_ANGLE} is a small angle. 

\begin{table}[ht]
    \centering
    \caption{LOGO Primitives}
    \vspace{0.1in}
    \begin{tabular}{p{3.5cm}|p{3.5cm}}
    \hline
    \textbf{Primitive} & \textbf{Description} \\
    \hline
    \sfname{forward(dist)} & move forward \sfname{dist} units \\
    \hline
    \sfname{left(theta)} & rotate left by \sfname{theta} degrees \\
    \hline
    \sfname{right(theta)} & rotate right by \sfname{theta} degrees \\
    \hline
    \sfname{penup\text{()}} & lift the pen (stop drawing) \\
    \hline
    \sfname{pendown\text{()}} & put the pen down (start drawing) \\
    \hline
    \sfname{teleport(x, y, theta)} & move to position \sfname{(x, y)} with angle \sfname{theta} \\
    \hline
    \sfname{heading\text{()}} & get the current angle of the turtle  \\
    \hline
    \sfname{isdown\text{()}} & check if the pen is down \\
    \hline
    \sfname{embed(program, vars)} & runs the code in \sfname{program} using the current context \sfname{vars} and teleports back to the original position.  \\
    \hline
    \end{tabular}
    \label{tab:logo_primitives}
\end{table}

\subsubsection{Date understanding} \label{append:date}
Date understanding involves both mathematical reasoning and parsing. 
Each question poses a word problem that requires reasoning about dates and times. 
For example, problems ask questions like: \emph{``On May 9th, 2017 Jane bought 40 eggs. She ate one per day. Today she ran out of eggs. What is the date 10 days ago in MM/DD/YYYY?''}. 
These types of questions are especially hard for LLMs to answer directly.
\citet{lyu.q.2023faithful} approached this task as a program prediction task, wherein an LLM predicts a Python script that gives the answer to the question when executed.
This paradigm is especially helpful for Date, as there are existing Python libraries that can perform math on dates, such as \sfname{datetime}() and \sfname{dateutil}(). 
While predicting programs with these libraries results in strong performance as compared to LLM-based reasoning, \citet{lyu.q.2023faithful} method predicts programs one at a time, leaving the benefits of shared sub-routines on the table. 
We use the programs predicted by \citet{lyu.q.2023faithful} as a starting point for our refactoring process.
\cref{tab:date_primitives} describes the Python libraries called by \citet{lyu.q.2023faithful}'s programs, which we treat as the primitives for Date. 
Date was released under an Apache 2.0 License.

\begin{table}[ht]
    \centering
    \caption{Date Primitives}
    \vspace{0.1in}
    \begin{tabular}{p{3.5cm}|p{3.5cm}}
    \hline
    \textbf{Primitive} & \textbf{Description} \\
    \hline
    \sfname{date\text{()}} & returns a \sfname{date} object \\
    \hline
    \sfname{time\text{()}} & returns a \sfname{time} object \\
    \hline
    \sfname{relativedelta(time)}  & performs addition/subtraction of \sfname{time}, which can be days, weeks, months, or years. \\
    \hline
    \sfname{strftime(format)} & prints the date in the specified format \\
    \hline
    \end{tabular}
    \label{tab:date_primitives}
\end{table}

\subsubsection{TextCraft} \label{append:textcraft}
TextCraft consists of goal queries paired with crafting recipes. Recipes are presented with distractors, making the parsing task challenging. 
Furthermore, the agent must reason about preconditions, as items can only be crafted when the requisite ingredients have been collected. 

The queries ask for a particular item to be crafted. 
For example the query can be ``\emph{craft behive}'' along with crafting commands:
\begin{quote}
    \textit{craft 4 oak planks using 1 oak log\\
    craft 1 honeycomb block using 4 honeycomb\\
    craft 1 beehive using 6 planks and 3 honeycombs\\
    craft 1 white bed using 3 white wool, 3 planks, 
    etc.}
\end{quote}
\noindent The environment has three primitives: \sfname{inventory}, \sfname{craft}, and \sfname{get} which we convert into Python variants (\cref{tab:textcraft_primitives}). This dataset uses the Apache 2.0 license. To obtain primitive programs in the train set, we use all the depth-2 instances not in the test set, and use the \textsc{ADaPT}~\citep{prasad2023adapt} trajectories with $d=4$. We then perform rule-based translation of environment commands into calls to primitive operations and filter out erroneous actions (based on the saved textual environment feedback) and all thought statements.

\begin{table}[ht]
    \centering
    \caption{TextCraft Primitives}
    \vspace{0.1in}
    \begin{tabular}{p{3.5cm}|p{3.5cm}}
    \hline
    \textbf{Primitive} & \textbf{Description} \\
    \hline
    \sfname{getObject(obj\_name)} & get \sfname{obj\_name} from the environment \\
    \hline
    \sfname{craftObject(obj\_name,} \sfname{\hspace{3em}[ingredients])} & craft \sfname{obj\_name} using the list of ingredients\\
    \hline
    \sfname{checkInventory\text{()}} & return the contents of the inventory \\
    \hline
    \end{tabular}
    \label{tab:textcraft_primitives}
\end{table}

\subsubsection{MATH}
MATH \citep{hendrycks2021measuring} contains challenging math word problems with open-ended answers (i.e. not multiple-choice). 
Unlike LOGO, Date, and TextCraft, MATH does not have any primitives, as each problem can be solved using standard Python math tools (addition, subtraction, etc.); this makes MATH a challenging setting for discovering helper functions, as the space of programs is less constrained. 

\subsubsection{TabMWP}
Like MATH, TabMWP \citep{lu2022dynamic} is a dataset of word problems. 
In this case, the word problems pertain to tabular data, where each problem asks questions involving computation over tabular data. 
This dataset has been used in prior work on tool learning, e.g. \citet{yuan.l.2023craft}. 
Like MATH, there is no existing domain-specific language for TabMWP.

\section{Analysis}

\subsection{What kinds of programs are discovered} \label{append:qual}
\cref{tab:logo_programs,tab:date_programs,tab:tc_programs} show examples of the discovered programs most commonly used by the agent. 

\begin{figure}[ht]
    \centering
    \scriptsize
\begin{python}
    def draw_small_5gon(): 
        for i in range(5):
            forward(2)
            left(72.0)
    def draw_semicircle():
        for i in range(HALF_INF):
            forward(EPS_DIST * 2): 
            left(EPS_ANGLE) 
\end{python}  
    \caption{Examples of discovered programs for LOGO as mentioned in \cref{fig:fig1,fig:usage}. As the name suggests, \sfname{draw\_small\_5gon}() draws a small-size pentagon and \sfname{draw\_semicircle}() draws a small semicircle.}
\label{tab:logo_programs}
\end{figure}

\begin{figure}[ht]
    \centering
    \scriptsize
    \begin{python}
    def get_date_today(date_obj):
        return date_obj
    def get_date_one_week_from_today(date_obj):
        return date_obj + relativedelta(weeks=1)    
    def get_date_one_week_ago(date_obj):
        return date_obj - relativedelta(weeks=1)
    def get_date_one_year_ago(date_today):
        return date_today - relativedetla(years=1) 
    \end{python}
    \caption{Examples of common discovered programs for Date as mentioned in \cref{fig:usage}. All the helpers functions and variables are named intuitively reflection their functionality.}
    \label{tab:date_programs}
\end{figure}
    
\begin{figure}[ht]
    \centering
    \scriptsize
    \begin{python}
    def craft_object_with_ingredients(target,
        ingredients):
        inventory = check_inventory()
        for ingredient in ingredients:
            if ingredient not in inventory:
                get_object_from_env(ingredient)
            craft_object(target, ingredients)
    def check_and_get_object(target):
        inventory = check_inventory() 
        if target not in inventory:
            get_object(target)
    \end{python}
    \caption{Examples of common discovered programs for TextCraft as shown in \cref{fig:fig1,fig:method}. \sfname{craft\_object\_with\_ingredients}() encapsulate the game dynamics of TextCraft as it first fetches the inventory, and every ingredient not in the inventory is obtained from the environment prior to using the crafting the target object. Similarly, the helper \sfname{check\_and\_get\_object}() gets an object from the environment if it is not already in the inventory. }
        \label{tab:tc_programs}

\end{figure}
    
\begin{figure}[ht]
    \centering
    \scriptsize
    \begin{python}
    def calculate_midpoint(point1, point2):
        x = (point1[0] + point2[0]) / 2
        y = (point1[1] + point2[1]) / 2
    return (x, y)

    def calculate_square(num):
        return num ** 2
    \end{python}
    \caption{Examples of discovered programs for MATH. \sfname{calculate\_midpoint} is an example of reuse, where \method{} finds a frequently-used functionality and encapsulates it, while in \sfname{calculate\_square} \method{} wraps a fairly easy function with a more informative name.}
        \label{tab:math_programs}

\end{figure}

\begin{figure}[ht]
    \centering
    \scriptsize
    \begin{python}
def calculate_total_cost(packages, items):

    total_cost = 0
    for item in items:
        total_cost += packages[item]
    return total_cost

def add_prices(prices):

    total_cost = 0
    for price in prices.values():
        total_cost += price
    return total_cost
    \end{python}
    \caption{Examples of discovered programs for TabMWP, reflecting the domain, which often involves costs and prices.}
        \label{tab:tabmwp_programs}

\end{figure}

\section{Hyperparameters} \label{append:hyperparams}

\cref{tab:hyperparams} lists the refactoring and testing hyperparameters used for each domain. 

\begin{figure}[ht]
    \centering
    \includegraphics[width=\linewidth]{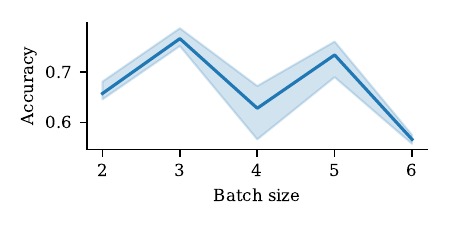}
    \vspace{-2em}
    \caption{Validation performance of CodeLlama-13B with \method{} abstractions trained using different batch sizes.}
    \label{fig:batchsize}
\end{figure}

\subsection{Varying Batch Size} \label{append:batch}
\cref{fig:batchsize} shows the performance of a CodeLlama-13B agent on the Date dev set for \method{} abstractions trained using different batch sizes. 
We see that performance can vary substantially and non-linearly with batch size, with 3 being the best setting and 5 a close second. 

\begin{table}[ht]
    \centering
    \caption{Hyperparameter settings for all experiments}
    \vspace{0.1in}
    \begin{tabular}{lccc}
    \hline
    \textbf{Setting} & \textbf{LOGO} & \textbf{Date} & \textbf{TextCraft} \\
    \hline
    Rounds of refactoring & 3 & 1 & 1 \\
    \sfname{editEvery} & 5 & 5 & 5  \\
    \sfname{pruneEvery} & 5 & 5  & 5 \\
    Add comments & True & False & False \\
    Batch size & 5 & 3 & 4 \\
    Filtering threshold & 0.0 & 0.0 & 0.0 \\
    Filter before testing & True & True & False \\
    ICL budget ratio & 0.5 & 0.5 & 0.5 \\
    \hline
    \end{tabular}
    \label{tab:hyperparams}
\end{table}

\section{Prompts} \label{append:prompts}
Below, we detail the prompts used in all sections. 
\begin{table*}[ht]
\caption{Batch refactoring prompt (\sfname{refactorBatch()}). \sfname{\/\/} Comments indicate where text is repeated. Brackets indicate variables filled in by the environment. Note that ``\sfname{<>} strings'' are passed as-is to the LLM. } 
\vspace{0.1in}

\label{prompt:refactor}
\centering
\begin{tabular}{p\linewidth}
\toprule
\begin{prompt}
Please rewrite the following two programs to be more efficient. 
{primitive description string}
The resulting programs MUST execute to the same result as the original programs.
Start by writing helper functions that can reduce the size of the code.  
You can also choose from the following helper functions:
{codebank function definitions} 

// repeated for all in batch
QUERY {i}: {query}
PROGRAM {i}: {program} 

Please format your answer as:
// repeated for i 
NEW PROGRAM {i}:
// once at end 
NEW HELPERS: 

Do not include any text that is not valid Python code.
Recall that no matter what, your program MUST be formatted in the following fashion: 
// repeated for i
NEW PROGRAM {i}:
# Thoughts:
# 1. The query asks for: <query intention> 
2. <query> can be solved by <components>.
# 3. I will use helper function <function> to <goal>.
<code for program {i}> 

Try to make your new programs as short as possible by introducing shared helper functions. Helper function parameters should be as general as possible and helper functions should be informatively named.
{logo_special_instr}

\end{prompt}\\
\hline
\sfname{logo\_special\_instructions} = 
\begin{prompt}
If the original function uses `embed`, you will likely need to use `embed` in your version. All code to be repeated needs to be included within the triple quotes passed to embed.
\end{prompt}\\
\hline
\end{tabular}
\end{table*} 

\begin{table*}[ht]
\caption{Query decomposition prompt. Output is used by the comment prompt in 
    \vspace{0.1in}
\cref{prompt:comment} } 
\label{prompt:decompose}
\centering
\begin{tabular}{p\linewidth}
\toprule
\begin{prompt}
You are an expert coder. For each query below, decompose it into its parts. 
Example: 
Query: Do some action 5 times and then do another action
Query (decomposed): 
The query asks: Do some action and then do another action
This can be decomposed into: 
1. repeat an action 
2. some action
3. another action

Query: {query}
Query (decomposed):
\end{prompt}\\
\hline
\end{tabular}
\end{table*} 

\begin{table*}[ht]
\caption{Prompt to add comments to primitive programs. Takes output of \cref{prompt:decompose} as input. } 
    \vspace{0.1in}

\label{prompt:comment}
\centering
\begin{tabular}{p\linewidth}
\toprule
\begin{prompt}
Please add comments to the following program to explain what each chunk of code does with respect to the query. 
First, decompose the query into parts. Then comment the code with the query parts. 
Example: 
Query: Do some action and then do another action
Code: 
do_some_action()
do_another_action()

Query: Do some action 5 times and then do another action
Query (decomposed): 
The query asks: Do some action and then do another action
This can be decomposed into: 
1. repeat an action 
2. some action
3. another action
Commented code:
# repeat an action
for i in range(5): 
    # do some action
    do_some_action()
# do another action
do_another_action()

{primitive description}

Query: {query}
Code:
{program}

Query (decomposed):
{output from decompose prompt}
\end{prompt}\\
\hline
\end{tabular}
\end{table*} 

\begin{table*}[ht]
\caption{Prompt for \sfname{editCodeBank}.} 
    \vspace{0.1in}

\label{prompt:edit}
\centering
\begin{tabular}{p\linewidth}
\toprule
\begin{prompt}
Refactor the following function to improve performance. 
FUNCTION: 
```
{func_str}
```

{library_str}

You may also use the following helper functions: 
{codebank_str} 

Try to increase the number of passing programs. Try to make programs general. For example, you can add parameters instead of hardcoded values or call other helper functions. First, for each failing query, explain why the programs do not accomplish the query's goal. Output this reasoning as: 
Thoughts:
1. The function passes some tests and fails others because <reason>. 
2. The failing queries <repeat queries here> asked for <intent>. 
3. The program failed because <reason>. 
4. This can be addressed by <change>. 
Then output your program so that all test cases pass, using the following format: NEW PROGRAM: <program>
Currently, {func._name} passes in {pass_perc * 100:.1f}\% of cases and fails in {fail_perc*100:.1f}\%.

SUCCEEDED:
{example of passing case}
FAILED: 
{example of failing case}
Thoughts:
\end{prompt}\\
\hline
\end{tabular}
\end{table*} 

\begin{table*}[ht]
\caption{Agent prompt for Python tasks like Date understanding. Note that the baseline and \method{} agent use the same prompt, but \sfname{\{codebank\_str\}} is empty for the baseline agent, and the \method{} sees some demonstrations from the Demo Bank in \sfname{\{icl\_string\}}.} 
    \vspace{0.1in}

\label{prompt:python_agent}
\centering
\begin{tabular}{p\linewidth}
\toprule
\begin{prompt}
Your task is to solve simple word problems by creating Python programs.
{codebank_str}

You will be given a query and have to produce a program. {thought_str} 
Examples:
{icl_string}

Please generate ONLY the code to produce the answer and nothing else.
Query: {query}
{thought_and}Program:
\end{prompt}\\
\hline
\end{tabular}
\end{table*}

\begin{table*}[ht]
\caption{Prompt for the LOGO agent. } 
    \vspace{0.1in}

\label{prompt:logo_agent}
\centering
\begin{tabular}{p\linewidth}
\toprule
\begin{prompt}

Your task is to draw simple figures using python Turtle graphics. 
You will use a custom turtle library, similar to the built-in library, which is sufficient for all tasks. 

Here's a description of the custom library: 
- forward(x): move forward x pixels
- left(theta): rotate left by theta degrees
- right(theta): rotate right by theta degrees
- penup(): stop drawing
- pendown(): start drawing
- teleport(x, y, theta): move to position (x, y) with angle theta
- heading(): get the current angle of the turtle 
- isdown(): check if the pen is down
- embed(program, local_vars): runs the code in program using the current context and teleports back to the original position. Allows you to nest programs. Implementationally, embed gets the turtle state (is_down, x, y, heading), executes program, then returns to the original state.
- save(path): save the picture to file 
{codebank_str}

You will be given a query and have to produce a program. Begin your program with a comment that explains your reasoning. For example, you might write:\n# Thought: the query asks for a line, so I will use the forward() function.
Examples:
{icl_string}

Please generate ONLY the code to produce the answer and nothing else.
Query: {query} 
Thought and Program: 

\end{prompt}\\
\hline
\end{tabular}
\end{table*}

\end{document}